\documentstyle[12pt,aps,psfig]{revtex}
\begin{document}
\title{ Cherenkov-drift emission mechanism }
\author{ Maxim Lyutikov}
\address{Theoretical Astrophysics, California Institute of Technology,
Pasadena, California 91125}
\author{ George Machabeli}
\address{ Abastumani Astrophysics Observatory, A. Kazbegi Av. 2a, Tbilisi, 
380060
 Republic of Georgia}
\author{ Roger Blandford}
\address{Theoretical Astrophysics, California Institute of Technology,
Pasadena, California 91125}

\date {1997}
\maketitle

\date { 1 April  1997}

\begin{abstract}
Emission of a charged particle  propagating in a medium with  a  curved magnetic  
field is reconsidered stressing the analogy between this emission 
mechanism  and collective  Cherenkov-type plasma emission.
It is explained how   this mechanism
differs from  conventional Cherenkov, cyclotron or curvature
emission and how it 
includes, to some extent, the features of each of these mechanisms.
Presence of  
a medium supporting subluminous waves  is essential for the possibility of wave
 amplification by 
 particles streaming along the curved magnetic field with a finite curvature
drift. We suggest an analogy between the curvature drift emission and the 
anomalous cyclotron-Cherenkov emission. Treating the 
emission in  cylindrical coordinates in the plane-wave-like approximation
allows one to compute the  single
particle emissivity and growth rate of the Cherenkov-drift instability.
 We compare the growth rates calculated using the single
particle emissivity and using  the dielectric tensor of one dimensional
plasma streaming  along the curved field. In calculating the single particle
emissivity it is essential to know the normal modes of the medium and their
polarization which can be found from  the dielectric tensor of the medium.
 This emission mechanism may be important for the problem of pulsar
radio emission generation.
% Cherenkov-drift  instability
%developing in pulsar magnetosphere is  considered including quasilinear stage. 
\end{abstract}

\section{Introduction}

Studies of relativistic  strongly magnetized plasma in 
astrophysical  setting (like pulsar magnetosphere) have shown the
 possibility of 
a new mechanism of electromagnetic wave generation. This new mechanism, which
combines features of conventional Cherenkov, cyclotron and curvature
radiation, deserves more detailed consideration from the  fundamental physics
point of view. Besides investigating the physical nature of this process, we
also  reconcile
different available approaches to this problem.

In this work we discuss this  novel emission mechanism 
of a charged particle streaming with relativistic velocity along curved
magnetic field line in a medium. A weak inhomogeneity of the magnetic field
results in a curvature  drift motion of the particle perpendicular to the 
local plane of the magnetic field line. A gradient drift 
(proportional to ${\bf \nabla \cdot B}$) is 
much smaller than the curvature drift and will be neglected. 
When the motion of the particle
parallel to the magnetic field is ultrarelativistic the drift motion even in the 
weakly inhomogeneous field can become weakly relativistic resulting in a new type
of generation of {\it electromagnetic}, vacuumlike waves. 
 Presence of three ingredients  ( strong but finite magnetic
field,  inhomogeneity of the field  and a medium with the index of refraction larger
than unity) is essential for the emission.  We will call this mechanism
Cherenkov-drift emission stressing the fact that microphysically it is virtually 
Cherenkov-type emission process.

Conventional consideration of the curvature emission (\cite{Blandford1975}, 
\cite{ZheleznyakovShaposhnikov}, 
\cite{MelroseLou},  \cite{Melrosebook1}) emphasize the 
analogy between  curvature emission and  conventional cyclotron emission. 
To our opinion this approach, though formally correct, has  limited 
applicability and  misses some  important 
physical properties of the emission mechanism. In a separate approach Kazbegi {\it et al.}
\cite{Kazbegi} considered this process calculating a dielectric
tensor of the inhomogeneous magnetized medium, thus treating the 
emission process as a collective effect. They showed that  maser 
action is possible only if a  medium supports subluminous waves.
In this work we show how these two approaches can be reconciled and
argue that the  dielectric
tensor  approach, which treats the Cherenkov-drift emission as a collective 
process, has a wider applicability.

%stress the fact that the curvature maser is possible only in a medium supporting 
%subluminous waves and thus can be considered as Cherenkov-type 
%instability (see below).  

The interplay between cyclotron
 (or synchrotron) and Cherenkov radiation has been
a long-standing matter of interest. Schwinger  {\it et al.}
 \cite{Schwinger} discussed the
relation between  these two seemingly different emission mechanisms.
They showed  that 
 conventional synchrotron emission
and Cherenkov radiation may be 
 regarded as respectively limiting cases
 of $|n-1|\, \ll \,1 $ and $B=\,0$
of a synergetic (using the terminology of Schwinger  {\it et al.}
\cite{Schwinger}) cyclotron-Cherenkov radiation.
In another work \cite{LyutikovCher} this analogy has been discussed  to include 
cyclotron-Cherenkov emission at the 
anomalous Doppler resonance. 
An important new aspect of our work (as compared with \cite{Schwinger} and
\cite{LyutikovCher} )
is that we take into account  inhomogeneity of the medium. 

%Generally,
%the synergetic  emission of electromagnetic waves by charged particles in a magnetized
%dielectric has the features of both  synchrotron and Cherenkov type emission.
%We will refer to the particular emission  process as a Cherenkov-type emission
%if  it is the Cherenkov part of the synergetic emission process that is essential for the
%emission of the waves. Alternatively, we will call the 
 %particular emission  process as a synchrotron-type emission if it is the 
%synchrotron part of the synergetic emission process that is essential for the
%emission of the waves. 
Physical origin of the emission in the case of Cherenkov-type and
synchrotron-type  processes is quite different. In the case of   Cherenkov-type
process  the emission may be  attributed to the electromagnetic 
polarization shock front that 
develops in a dielectric medium due  to the passage of a charged particle with  
speed larger than phase speed of  waves in a medium. It is virtually a collective
emission process.  In the case of  
synchrotron-type  process, the emission may be attributed to the Lorentz force acting 
on a particle in a magnetic field.  Cherenkov-type
emission is impossible in vacuum and in a medium with the refractive index smaller than
unity. 

Cyclotron emission at the anomalous Doppler effect (cyclotron-Cherenkov emission) 
 is an interesting example of 
Cherenkov-type emission process of a particle in a magnetic field. It is impossible in vacuum
and requires a superluminous motion of a particle along the magnetic field. Thus, the
emission
at the  anomalous Doppler effect is 
 attributed to the 
polarization shock front that a spiraling particle induces in a medium.
In our opinion, the 
Curvature-drift emission may be viewed as  a Cherenkov-type 
emission that bears the same relation to the conventional curvature emission as the 
cyclotron emission at the anomalous Doppler effect bear to the conventional 
cyclotron emission.

In this work we consider a 
Curvature-drift emission   of the particles in the ground
gyrational state. 
It is possible  to obtain the emissivity for particles in  
excited gyrational state by the method of dielectric tensor 
\cite{Akhalkatsi1997}. Then,  conventional  cyclotron (cyclotron-Cherenkov),
 Cherenkov and
curvature emission mechanisms
 may be viewed as corresponding limits of the Cherenkov-drift 
mechanism in the cases of homogeneous magnetic
field (in  a medium), medium without magnetic field, and inhomogeneous 
magnetic field without a medium.  

In  Section \ref{Model1} we discuss our set up of the problem and how it differs
from the previous consideration. In Section \ref{emissivity}
we calculate a single particle emissivity of the Cherenkov-drift 
mechanism  and find the growth for  kinetic beam instability toward excitation
of electromagnetic waves at the Cherenkov-drift resonance. Then the results are
compared with those obtained by the dielectric tensor method. 
%And finally in Section \ref{Applcation} we discuss shortly the possible application
%to the problem of pulsar radio emission.

\section{Description of a model}
\label{Model1}

\begin{figure}
\psfig{file=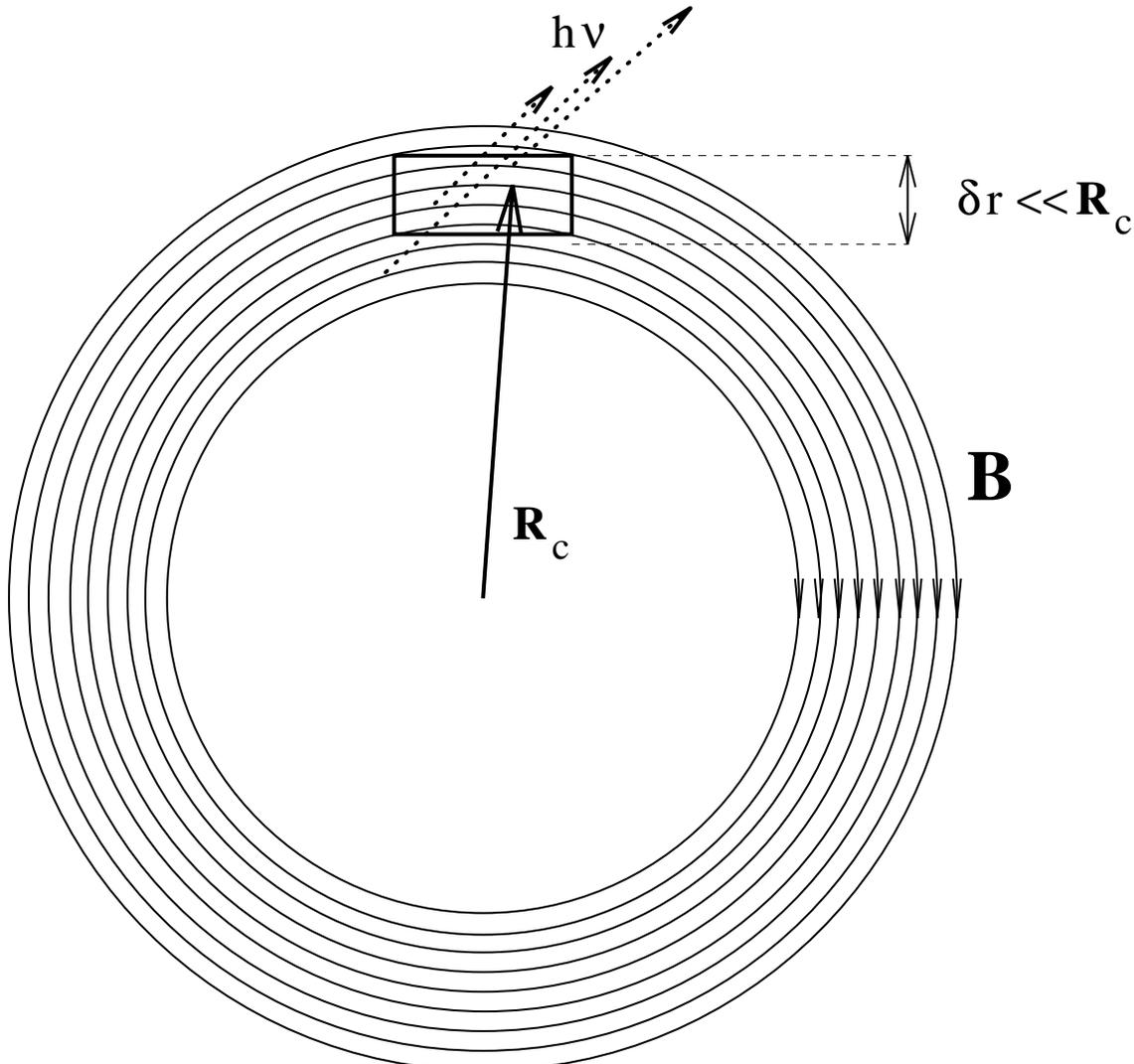,width=15.0cm}
\caption{ 
Geometry of  the considered problem. The magnetic field lines are
 concentric  coplanar
 circular arcs with the radius of
 curvature much larger than the size of the region.
\label{max1}}
\end{figure}

Let us consider concentric  coplanar
 circular magnetic field lines
populated with charged 
particles streaming relativistically
 along the curved  magnetic field (Fig. \ref{max1}). 
The  magnetic field
is assumed to be very strong, so that the particles are 
in their ground gyrational state (zeroth Landau level).  Next we 
introduce a cylindrical system coordinates $r, \phi, x$
 with the unit vector  
 ${\bf e}_{\phi} $ along the local magnetic field,
${\bf e} _x$ perpendicular to the osculating plane of the magnetic field and 
${\bf e_r} = 
{\bf e}_{\phi}
\times {\bf e} _x$. 
We will limit the region under consideration to have the 
size along the  $r$ axis $\delta r$ much smaller than the typical 
 radius of curvature of the field lines in the region
($\delta r \ll R_B$) and much larger than the typical wave length 
of the considered waves
($\delta r \gg  \lambda $).
 %With the density of the particles and their velocities constant 
%across the region,   the system considered is homogeneous  
%anisotropic medium in a strong
%circular magnetic field, e.i. magnetic field is constant in absolute value but changes its
%direction from point to point. 
This  physical picture may be considered as an approximation to 
the region of the open field 
lines of the pulsar magnetosphere. The curved dipole field   
can be locally
approximated by the circular field with the radius of curvature  $R_B $. 
The 
typical radius of curvature $ R_B \approx 10^9 $ cm and the wave length of the emitted radiation 
is of the order of 
meters $\lambda \approx 10^2 $ cm. Then for the size of the 
region satisfying the
 condition $  10^2 {\rm cm} \ll \delta r  \ll 10^ 9$ cm,  it can be considered
as  homogeneous. 

Next, we assume,  that the collective interactions of the particles with  waves 
propagating in the system may be described using a
dielectric tensor. 
Then, for a given wave mode, all the particles may be separated in the 
plasma particles, i.e.
 those  particles   that interact with the 
wave nonresonantly,  and resonant 
particles. 
We expect that outgoing electromagnetic modes generated by some
fluctuating currents at smaller radii
will interact with resonant particle and will
be amplified as they propagate through the interaction region
(see Fig. \ref{max1}).

%Due to the cylindrical symmetry we can then identify the flux observed
% at infinity through some angle $d \theta$ as been emitted by the
%particles inside a corresponding angle $d  \phi$. 

We wish to find the  specific intensities for emission of electromagnetic waves 
  by the  resonant particles in
 such a system. This may be done by calculating the work done by the extraneous
current associated with the resonant particle moving along a given trajectory 
(\cite{Melrosebook1})
 or by calculating the dielectric tensor of the medium.  Below
we discuss  in details the calculations using the former method and 
compare it with the
results obtained using the dielectric tensor.

In a curved magnetic field 
  a particle with  relativistic Lorentz factor $\gamma$ streaming 
along the field line  experiences a drift in the $x$ direction with 
velocity 
\begin{equation}
u_d = { \gamma _{\phi} v_{\phi}^2 \over \omega_B R_B } ,
\label{qaq}
\end{equation}
where $\omega_B=\, q \, B/(m\, c)$  is the 
nonrelativistic gyrofrequency, $q$ is charge of the particle, $m$ is its mass and $c$
is the speed of light.

With the instantaneous
  3-velocity  ${\bf v } = \{ 0, v_{\phi}, u_d\}$ the  radius vector of the particle
is 
\begin{equation}
{\bf r} (t)  =  \{ R_B, v_{\phi} t , u_d t \} .
\label{qaq01}
\end{equation}
The current density associated with the charge is 
\begin{equation}
{\bf j} ({\bf r}  , t)   =  q {\bf v }  \delta ( {\bf r}  - {\bf r} (t) ).
\label{qaq02} 
\end{equation}
We next find the Fourier transform of the current
  (\ref{qaq02}) in  cylindrical 
system of coordinates by 
expanding the current density in real space  (\ref{qaq02}) 
 in terms of cylindrical waves $ \exp \left(-
  i ( k_r r + k_x x + m \phi) \right)$  (m - integer) :
\begin{equation}
{\bf j} (\omega, {\bf k} )  =
\int _{ - \infty} ^{\infty} d t   \int d  {\bf r} \,
{\bf j} ({\bf r}  , t) \exp\left( i ( \omega t - k_r r - k_x x -  m \phi )\right) .
\label{qaq04}
\end{equation}

Using the relation 
\begin{equation}
\int _{ - \infty} ^{\infty} d t \exp  \left(  i \alpha t \right) = 2 \pi \delta ( \alpha) 
\label{qaq03}
\end{equation}
 and the radius vector (\ref{qaq01})
 we find the 
Fourier image of the current:
\begin{equation}
{\bf j} (\omega, {\bf k} ) =  2 \pi q {\bf v }  \exp \{ - i k_r R_B\} 
\delta ( \omega - k_{\phi} v_{\phi} - k_x u_d),
\label{qaq1}
\end{equation}
where we introduced $ k_{\phi} = m/R_B $. For $m \gg 1$ we can assume that 
$ k_{\phi} $ is continuous. 

The condition, that the size of the region considered is much larger than the 
wave length places a lower limit on the wave frequency in (\ref{qaq1}):
\begin{equation}
 \omega  \gg \Omega ,
\label{qaq11}
\end{equation}
where $\Omega = v_{\phi} /R_B$ is the angular frequency of the particle's
 rotation along the circular
magnetic field. The condition (\ref{qaq11})
also vindicates the assumption $ m \gg 1$.

The expansion (\ref{qaq04}) in cylindrical coordinates 
has a limited applicability \cite{LyutikovMachabeliBlandford2}.
Generally, in cylindrical coordinates the normal modes will be expressed in terms
of Bessel-type functions.
For the nonresonant  modes we can use WKB approximation to the radial dependence
of  normal modes (this is equivalent to the tangent expansion of Bessel functions
when argument is larger and not close to the order). 
On the other hand,
for the resonant modes the  argument of Bessel functions
is close to the order, so that the  WKB approximation (or expansion
in tangents) is not applicable. In this case we can use Airy function
approximation to Bessel function, which, in turn, has a wave-like approximation 
for the interaction of {\it subluminous} waves with the  particles
moving  with the speed larger than the  speed of light in a medium 
(this corresponds to the Airy function expansion
argument is larger than the order).  It is shown in \cite{LyutikovMachabeliBlandford2}
that for the large argument expansionof Airy  function to apply
it is required that (i) $ n -1 \gg 1/\gamma^2$ and (ii) $ (  n -1) m ^{2/3} \geq 1$
(here $n$ is the refractive index, $\gamma$ is the Lorentz factor of the
resonant particle. 
Summarizing the above: 
expansion (\ref{qaq04}) is valid in the two different regimes:
(i) for the  nonresonant  modes, (ii) for the resonant modes
when $ n -1 \gg 1/\gamma^2$ and $ (  n -1) m ^{2/3} \geq 1$.
In what follows we assume that the conditions for the expansion (\ref{qaq04})
are satisfied.

An important difference of this approach from  the one used in 
\cite{Blandford1975}
 and 
\cite{MelroseLou} is that  we  calculate the transition current in cylindrical coordinates,
 while in  \cite{Blandford1975}, 
 and 
\cite{MelroseLou} the transition current was calculated in 
Cartesian coordinates. In adopting  Cartesian coordinates with planar
normal modes  the interaction length for an individual
electron,  $\approx R_B /\gamma_b$,
was essentially coextensive with region over which the waves could
possibly interact with any electron.  This introduces a strict constraint 
on the particle-wave interaction and  precluded a  strong amplification
under all circumstances because the wave would have to grow substantially
during a very short interaction in a manner that could not be easily
quantified. The second major difference from \cite{Blandford1975} and
\cite{MelroseLou} is that we consistently take dispersion into account.

The major  advantage of cylindrical system of coordinates
 is that  the only inhomogeneity present in the problem,
the weak inhomogeneity  of the direction of magnetic field,  
can be effectively eliminated by transforming to the cylindrical system of coordinates.
It is the choice of 
cylindrical coordinates that allows  one to  describe the weakly inhomogeneous
system by the homogeneous (independent of  {\bf r } )  set of equation,
but in the curved coordinates. 
The use of cylindrical waves with the phase dependence of the form
$ \exp \left(  i ( \omega t -  k_r r -  k_x x - m \phi) \right)$.
has a limited applicability, but allows for simple estimate of the 
Cherenkov-drift emissivity and growth rate.

\section{Cherenkov-drift emissivity and amplification}
\label{emissivity}

In this section we calculate the growth rate of the Cherenkov-drift 
instability by finding the Cherenkov-drift 
emissivity for the single particles on the zeroth Landau level streaming along
the curved magnetic field. We then compare it with the  calculations done 
using the dielectric tensor of a plasma in a curved magnetic field. The results of both 
calculations, obviously, coincide.

With the  known current density,
 the energy radiated by the particle into a given wave mode
$\alpha$ is given by (\cite{Melrosebook1}, Eq. (3.18))
\begin{equation}
U^{\alpha} = 4 \pi R_E^{\alpha} ( {\bf k}) 
 \left| {\bf e}^{ \ast \alpha} \cdot {\bf j} (\omega ^{\alpha}  , {\bf k} ) \right|^2 ,
 \label{qaq101}
\end{equation}
where $ {\bf e}^{ \alpha} $ is the polarization vector of the emitted mode
and $R_E^{\alpha}$ is the ratio of electric to total energy
in the wave, as defined by \cite{Melrosebook1}. 

To find the energy radiated  one has first to determine the normal
modes of the medium and find their polarization. 
In contract to the vacuum case, where one is 
free to chose arbitrary combination of  plane  transverse electromagnetic 
waves as normal modes,
in a medium
 the normal modes
 must be the eigenvectors of the corresponding dispersion equation. 
%A failure to chose  correctly the normal
% modes and their polarization has lead to the wrong emissivities and corresponding
% growth rates in the past.

 To determine the  the normal
modes and their polarization we must solve the dispersion equation
\begin{equation}
Det\left| \Lambda_{lm} \right|=0,
\hskip .2 truein \Lambda_{lm} =
 (k^2 \delta _{lm}-k_l k_m) {c^2\over \omega^2}-\epsilon_{lm} ,
\label{det}
\end{equation}
where $\epsilon_{lm} $ is the dielectric tensor of the medium.
 Dielectric tensor for a one dimensional plasma streaming 
along the strong magnetic field has been  calculated in  \cite{Kazbegi},
\cite{LyutikovMachabeliBlandford2}. 

Dispersion equation (\ref{det}) is very complicated - all the nine components of the 
matrix $\Lambda_{lm} $ are nonzero.
This is  different from the case of 
a plasma in a  
homogeneous magnetic field, where,
 due to the cylindrical symmetry around magnetic field,
some components  of $\Lambda_{lm} $ 
could be set to  zero without a loss of generality.

\subsection{Polarization of Waves in Anisatropic Dielectric}
\label{pl}

Below
we will restrict our consideration to the case of electron-positron plasma with the 
same distribution functions streaming along a 
superstrong magnetic field. 

\subsubsection{Infinitely Strong Magnetic Field}
\label{pl1}

In the infinitely strong magnetic field the dielectric tensor
is
\begin{equation}
\epsilon_{ij} =
\left(
\begin{array}{ccc}
1&0&0 \\
0&1-K&0 \\
0&0& 1
\end{array} \right)
\label{D043}
\end{equation}
where
\begin{equation}
K= { 4 \pi q^2 \over m_e } \int { d p_{\phi} \over \gamma^3}
{f (p_{\phi})  \over ( \omega - \Omega m) ^2 } =
{ 4 \pi q^2 \over \omega }  \int d p_{\phi} { v_{\phi} \over
\omega - \Omega m} { \partial f ^{(0)} \over \partial p_{\phi}}
\label{D044}
\end{equation}
$ f (p_{\phi})$ is a one dimensional  distribution  function.

\begin{figure}
\psfig{file=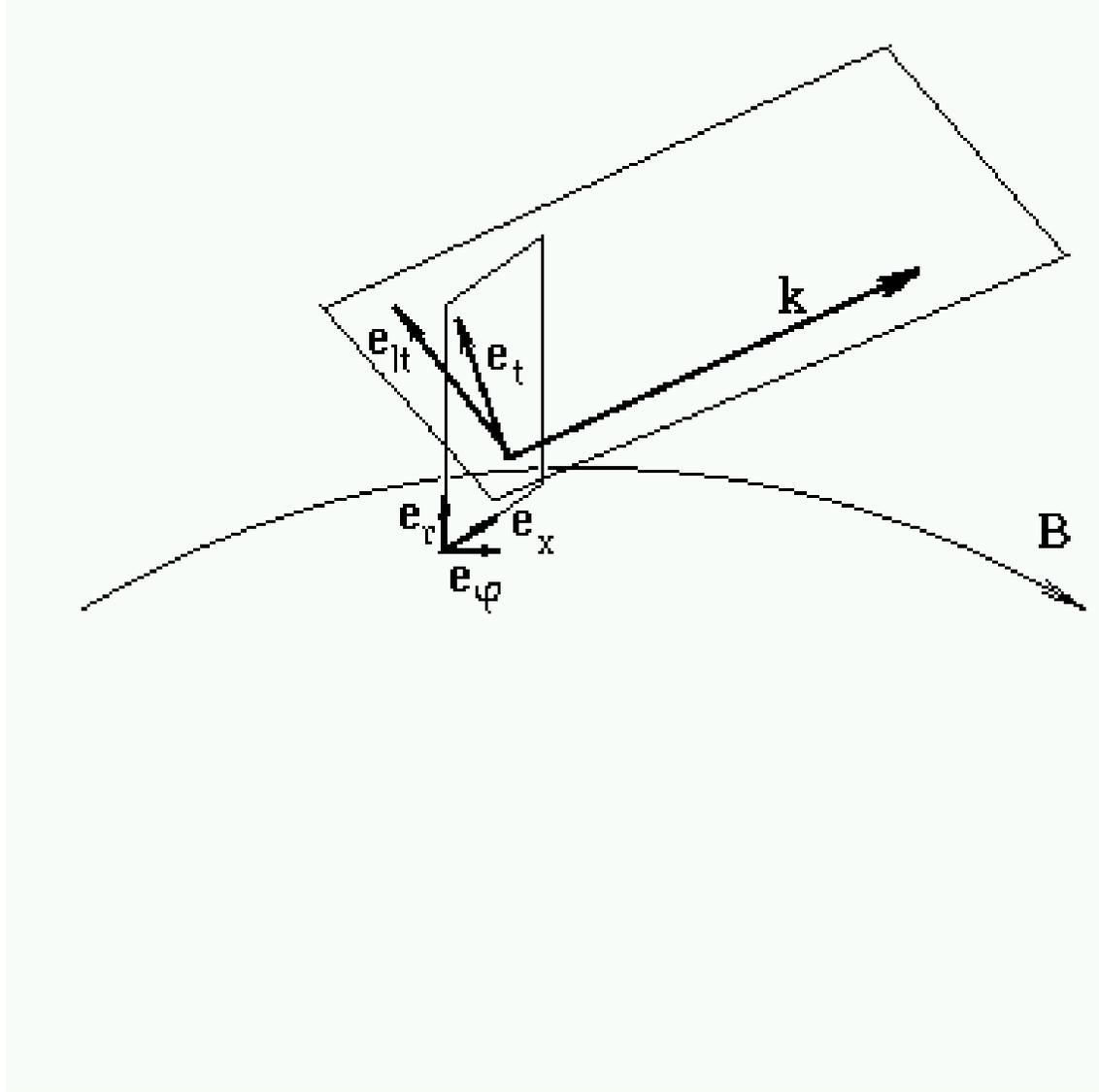,width=15.0cm}
\caption{ 
Normal modes in the considered
 medium in the limit of very strong magnetic field.
The electric field vector of the $t$-mode is in the plane
 ${\bf e_r} - {\bf e_x}$ and
the electric vector of the $lt $-mode is orthogonal to ${\bf e}_{t}-{\bf k}$ 
plane.
\label{Polariz}}
\end{figure}

The 
dispersion equation (\ref{D044}) may be factored for the dispersion relations
of 
 the t-mode, with the electric field perpendicular to the {\bf k} - {\bf B} plane, 
and lt-mode, with the electric field in the {\bf k} - {\bf B} plane (Fig. \ref{Polariz}):
\begin{eqnarray}
&&
n^2 =1 , \hskip 1 truein
 {\bf e}^{(t)} = { 1 \over n_{\perp}} \{ - n_x, 0, n_r \} 
\mbox{} \nonumber \\ \mbox{}
&&   n_{\phi}^2 = 1- { n_{\perp}^2 \over 1- K} ,
\mbox{} \nonumber \\ \mbox{}
&& {\bf e}^{(lt)} =
{1 \over \sqrt { ( 1-K)^2 + K n_{\perp}^2} }
\{ \sqrt { ( 1-K) ( 1-K - n_{\perp}^2) } { n_r \over n_{\perp}}, - n_{\perp},
 \sqrt { ( 1-K) ( 1-K - n_{\perp}^2) }
{ n_x  \over n_{\perp}} \}
\label{D51}
\end{eqnarray}
 
These waves are natural analogs of the t- and lt-modes in the case of 
straight magnetic field lines. The t-mode
is purely transverse and the lt-mode is a mixed transverse-longitudinal wave.
Generally, in the  lt-mode the electric field is not perpendicular to the 
wave vector, but for teneous plasma in the high frequency limit 
 $ K \approx 0$, so that 
 the lt-wave is approximately transverse:
\begin{equation}
{\bf e}^{(lt)} =
{ 1\over n_{\perp}} \{ - n_r n_{\phi} , n_{\perp} , -n_x  - n_{\phi} \}
\label{D511}
\end{equation}
 
\subsubsection{Finite Magnetic Field}
\label{pl2}
In the finite magnetic field the dielectric tensor is
\begin{equation}
\epsilon_{ij} =
\left(
\begin{array}{ccc}
1+2 d &0&0 \\
0&1-K&0 \\
0&0& 1+ 2 d
\end{array} \right)
\label{D043a}
\end{equation}
where  for cold plasma $ d=  \omega_p^2 /\omega_B^2$. Eq. (\ref{D53}).
Dispersion relation is then
\begin{equation}
\left(
- ( 1 - K ) n_{\phi}^2 + (1 + 2  d)  ( 1  - K  - n_r^2 -n_x^2) \right)
( 1 + 2 d - n_{\phi}^2 - n_r^2 - n_x^2) =0
\label{D53}
\end{equation}
which has  solutions
\begin{eqnarray}
&&
n^2 =(1+ 2 d), \hskip 1 truein
 {\bf e}^{(t)} = { 1 \over n_{\perp}} \{ - n_x, 0, n_r \}
\mbox{} \nonumber \\ \mbox{}
&&   n_{\phi}^2 = (1+ 2 d) \left( 1 - { n_{\perp}^2 \over 1- K} \right),
\mbox{} \nonumber \\ \mbox{}
&& {\bf e}^{(lt)} \approx
 {1 \over \sqrt { ( 1-K)^2 + K n_{\perp}^2}}
\{ \sqrt { ( 1-K) ( 1-K - n_{\perp}^2) } { n_x \over n_{\perp}}
, - n_{\perp},
{ n_r  \over n_{\perp}} \}
+ O( d)
\label{D54}
\end{eqnarray}
So that the polarization vectors are the same as in the case of infinitely
strong magnetic field within factors $ \omega_p^2 / \omega_B^2$.

\subsection{Cherenkov-drift emissivity}
\label{emissivity1}

With the 
${\bf k}$ vector of the emitted {\it transverse } waves given by
\begin{equation}
{\bf k} =\left\{ k_r, k_{\phi} , k_x \right\} ,
\label{qaq22}
\end{equation}
we  
 choose the following  polarization vectors 
\begin{eqnarray}
&&
{\bf e^{lt} } = {1\over k k_{\perp}   } \left\{ k_{\phi}  k_r, - k_{\perp}  ^2 ,
k_x k_{\phi} \right\},
\mbox{} \nonumber \\ \mbox{}
&&
{\bf e^t} ={1\over  k_{\perp}   } \left\{ - k_x ,0,  k_r, 
\right\},
\label{qaq23}
\end{eqnarray}
where $k_{\perp}  = \sqrt{k_r^2 + k_x^2 } $. 
This choice of polarization vectors is a limiting case of very strong magnetic field
and tenous plasma. 

We note, that 
the separation of the normal modes done in \cite{Blandford1975}, \cite{MelroseLou},  
\cite{Melrosebook1} is related 
 not to plane of the real magnetic field, directed along $\phi$,
and of the  vector ${\bf k}$,  but to the plane ${\bf k} - {\bf e_x}$.
This is justified only in vacuum or in the homogeneous medium where there 
a freedom in the choice of polarization vectors of the normal modes.
In an anizatropic  medium the normal modes and their polarizations have to be 
determined from the dispersion equation.

The eigenvectors (\ref{qaq23})
 are {\it different } from those chosen by 
\cite{MelroseLou},  \cite{Melrosebook1}. The modes chosen 
in those works follow the analogy between the curvature and synchrotron 
emission. They are {\it not the normal modes of the medium}. The approach of 
\cite{MelroseLou},  \cite{Melrosebook1}, involving vacuum wave
polarization and  refractive index close but not equal to unity,  
may be considered as
correction to the vacuum curvature emissivity due to  presence of a
medium when  effects of the medium on  wave propagation are small. 
When effects of the medium on  wave propagation  cannot be considered 
as  small perturbations one has to solve the dispersion relation to find the 
normal modes and their polarization. 
This stresses once again the 
importance of a medium in what we call the Cherenkov-drift emission. 
Presence 
of a medium changes the nature of the emitted modes and changes the 
corresponding emissivities.

With the polarization vectors (\ref{qaq23}),
the single particle probability of 
emission (per unit volume $d {\bf k} /( 2 \pi)^3$  may be written as a 
polarization tensor 
\begin{eqnarray}
&&
w^{\alpha \beta} ( {\bf k}, {\bf p}) 
={  4 \pi^2 q^2 \over \hbar \omega( {\bf k}) } 
F^{\alpha \beta} ( {\bf k}, {\bf p})  \delta( \omega - k_{\phi} v_{\phi}-
k_x u_d ) ,
\mbox{} \nonumber \\ \mbox{}
&&
F^{\alpha \beta}( {\bf k}, {\bf p})  = \left( \begin{array}{cc}
A^2&  i C A \\
- i C A & C^2 
\end{array} \right) ,
\mbox{} \nonumber \\ \mbox{}
&&
A={1\over k k_{\perp}  } \left(   u_d  k_{\phi} k_x v_{\phi} - k_{\perp}^2 v_{\phi}   \right),
 \mbox{} \nonumber \\ \mbox{}
&&
C ={  u_d  k_r \over k_{\perp}} ,
\label{qaq24}
\end{eqnarray}
with $\alpha, \, \beta = t$, $lt$. 
This form of the emissivity may be compared with \cite{Melrosebook1},
 Eq. (13.62-13.65),
 and 
\cite{MelroseLou}. There are two main differences: 
(i) the {\it approximate } expressions  in  \cite{Melrosebook1} and
\cite{MelroseLou} 
 for the 
single particle emissivity per unit frequency, involving Airy functions, 
are  obtained if the transition current is calculated in  Cartesian coordinates,
 while
relations (\ref{qaq24}) are exact, (ii) the polarization of the normal
modes chosen in \cite{Melrosebook1} and \cite{MelroseLou} are different from ours.

The total emissivity, summed over polarizations, follows from (\ref{qaq24}):
\begin{equation}
w ( {\bf k}, {\bf p}) 
={  4 \pi^2 q^2 \over \hbar \omega( {\bf k}) } 
\left[ \left( { k_{\phi} k_x  u_d  \over k k_{\perp} } - { k_{\perp} v_{\phi} \over k}  \right)^2 +
{ k_r^2 u_d  ^2 \over  k_{\perp} ^2 }  \right]
\delta( \omega - k_{\phi} v_{\phi} -
k_x u_d ) .
\label{qaq25}
\end{equation}
The first term in square brackets corresponds to the emission of the lt-mode,
second - to the emission of the t-mode. 

Next we calculate the growth rate of the Cherenkov drift instability (\cite{Melrosebook1}):
\begin{equation}
\Gamma = \int d {\bf p}  w ( {\bf k} ,  {\bf p} ) \, \hbar  {\bf k}  \cdot 
{\partial f ( {\bf p} ) \over \partial  {\bf p}  } .
\label{qaq3}
\end{equation}
Growth rate for the lt-mode is
\begin{equation}
\Gamma^{lt} = { 4 \pi^2 q^2\over m } \int d p_{\phi} 
\left( {k_{\phi} k_x u_d  \over c k k_{\perp}} - { v_{\phi} \over c} { k_{\perp} \over k} \right)^2
 { \partial f(p_{\phi}) \over  \partial p_{\phi} } 
\delta\left( \omega- k_{\phi} v_{\phi} - k_x u_d \right),
\label{qaq7}
\end{equation}
and  growth rate for the t-mode is
\begin{equation}
\Gamma^{t} = { 4 \pi^2 q^2\over m } \int d p_{\phi} 
\left( {k_r u_d  \over k_{\perp} c } \right)^2
 { \partial f(p_{\phi}) \over  \partial p_{\phi} } 
\delta\left( \omega- k_{\phi} v_{\phi} - k_x u_d \right).
\label{qaq71}
\end{equation}

Presence of delta function with the Cherenkov-drift  resonance
condition without the gyromagnetic term 
 indicates that this is a Cherenkov-type emission process which
requires that the medium supports  subluminous waves.
The necessary condition for the instability is also the same as in the    
 conventional Cherenkov instability: the derivative of the 
distribution function must be positive at the resonant frequency and
wave vector. In physical terms this means that the number of particles
with  the velocity larger than the phase velocity of the waves 
exceeds the number of particles with the velocity 
smaller than the phase velocity of the waves. This once again stresses the
Cherenkov-type nature of the emission. 

It is clear from (\ref{qaq7}) 
and (\ref{qaq71}) that the growth rate of the t-wave
is proportional to the drift velocity and becomes zero in the limit
of vanishing drift velocity.
  As for the lt-wave,  it can be excited in the limit
of vanishing drift by the conventional Cherenkov mechanism which does not
rely on the
  curvature of the magnetic field lines. We recall, that in the limit of 
a strong magnetic field  and {\it oblique} propagation
 lt-wave has two branches: one superluminous and
one subluminous (\cite{arons1}, \cite{LyutikovMachabeliUsov}). 
On the conventional Cherenkov  resonance  it is 
possible to excite only subluminous waves.
  We note that  the choice of polarization
vectors of \cite{Blandford1975},
 \cite{Melrosebook1} and \cite{MelroseLou}  excludes the 
excitation of the subluminous branch  as well
(for which electric field is not perpendicular to the {\bf k} vector)
thus prohibiting any maser action without taking into account drift motion.  

The growth rates (\ref{qaq71})
 and (\ref{qaq7})  may be compared with the calculation
done using the antihermitian part of the dielectric tensor \cite{Kazbegi} .
In case of kinetic
 instability the growth rate is  given by (\cite{Melrosebook1})
\begin{equation}
\Gamma = \left.- { (e_{\alpha}^{\ast} \epsilon^{\prime \prime}_{\alpha \beta}
 e_{\beta}) \over {1\over \omega^2 } { \partial \over \partial \omega }
\omega^2 (e_{\alpha}^{\ast} \epsilon^{\prime}_{\alpha \beta}
 e_{\beta}) } \right|_ { \omega= \omega({\bf k})},
\label{piajd}
\end{equation}
where $ \epsilon^{\prime}_{\alpha \beta}$ and $
\epsilon^{\prime \prime}_{\alpha \beta}$ are hermitian and antihermitian
parts of the dielectric tensor, $ \omega({\bf k})$  is the frequency 
of the excited normal mode of the medium,
and ${\bf  e}$ is its polarization vector.

The relevant components of the  antihermitian
part of the   dielectric tensor follow from \cite{Kazbegi},
\cite{LyutikovMachabeliUsov}:
\begin{eqnarray}
\epsilon^{\prime \prime}_{xx}&&= - i { 4 \pi^2 q^2 \over  \omega c }
\int {dp_{\phi} }   u_d^2 {\partial  f(p_{\phi} )  \over \partial p_{\phi} } 
\delta\left( \omega- k_{\phi} v_{\phi} - k_x u_d \right),
\mbox{} \nonumber \\ \mbox{}
\epsilon^{\prime \prime}_{x \phi} &&   = - i { 4 \pi^2 q^2 \over  \omega c }
\int {dp_{\phi} }  u_d v_{\phi} {\partial  f(p_{\phi} )  \over \partial p_{\phi} } 
\delta\left( \omega- k_{\phi} v_{\phi} - k_x u_d \right)
= \epsilon^{\prime \prime}_{\phi x} ,
\mbox{} \nonumber \\ \mbox{}
\epsilon^{\prime \prime}_{\phi \phi} &&   = - i { 4 \pi^2 q^2 \over  \omega  }
\int {dp_{\phi} }  v_{\phi} 
{\partial  f(p_{\phi} )  \over \partial p_{\phi} } 
\delta\left( \omega- k_{\phi} v_{\phi} - k_x u_d \right).
\label{qaq31}
\end{eqnarray}
Using (\ref{piajd}) and 
 (\ref{qaq31})  we confirm the growth rates (\ref{qaq71}) and (\ref{qaq7})
for $ v_{\phi} \approx c$.

This approach, involving  dielectric tensor of the medium may be considered
as a more general, than the one using the single particle
emissivities. The dielectric tensor approach takes consistently 
into account both resonant and nonresonant particles.  In calculating the 
single particle
emissivities one still  has to 
 calculate
the dielectric tensor to find the properties of the emitted normal modes of the medium. 

Next we estimate the growth rate for the Cherenkov-drift excitation
of electromagnetic waves in the strongly magnetized electron-positron 
plasma. 
In the plasma frame the dispersion relations for the transverse modes 
 in the limit $\omega - k_{\phi} v_{\phi} \ll \omega_B$ 
for quasi-parallel propagation ($k_r, \, k_x \ll k_{\phi}$) is 
\begin{equation}
 \omega = k c \, ( 1- \delta), \hskip .2 truein
\delta = {\omega^2 T_p \over \omega_B^2}\, \ll 1,
\label{det3}
\end{equation}
(\cite{Kazbegi}), where $\omega^2 = 4 \pi q^2 n_p/m$ is the plasma 
frequency of electrons or positrons, 
$ n_p $ -density of  plasma, $T_P$ is  effective temperature of 
plasma in units of $m c^2$. 

%  ??? CALCULATE THE DISPERSION OF THE LT MODE IN MODE DETAILS
% WHEN BACK IN LA ??????

\begin{figure}
\psfig{file=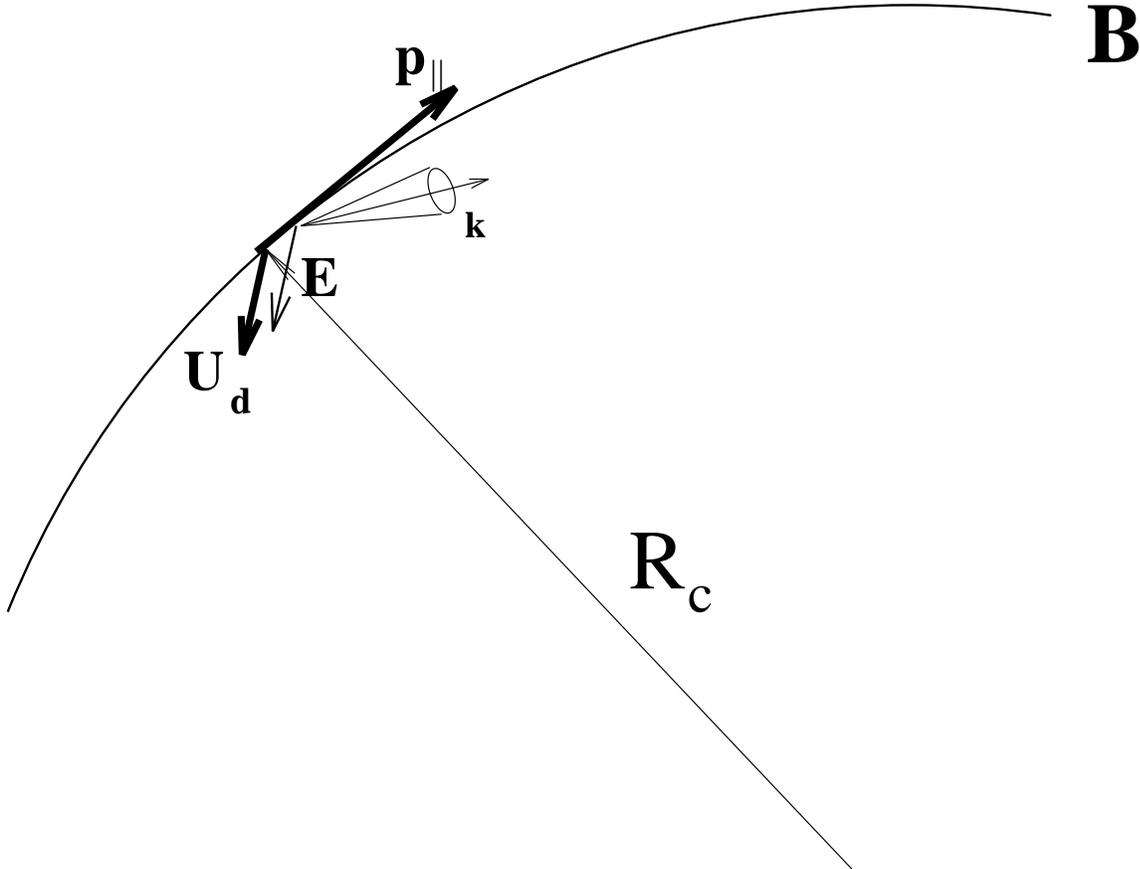,width=15.0cm}
\caption{
Cherenkov-drift emission in the case $\delta \ll u_d^2/c^2$.
 Drift velocity ${\bf u}_d$ is perpendicular
to the plane of the curved field line $({\bf B}-{\bf R_c} $
plane, ${\bf R_c} $ is a local radius of curvature).
The emitted electromagnetic waves
 are polarized along  ${\bf u}_d$. The emission is generated in the
 cone centered at the angle $\theta^{em} = u_d / c $ and
 with the opening angle  $(2 \delta)^{1/2} \ll \theta^{em}$.
\label{Cherenkov-driftemission}
}
\end{figure}

\begin{figure}
\psfig{file=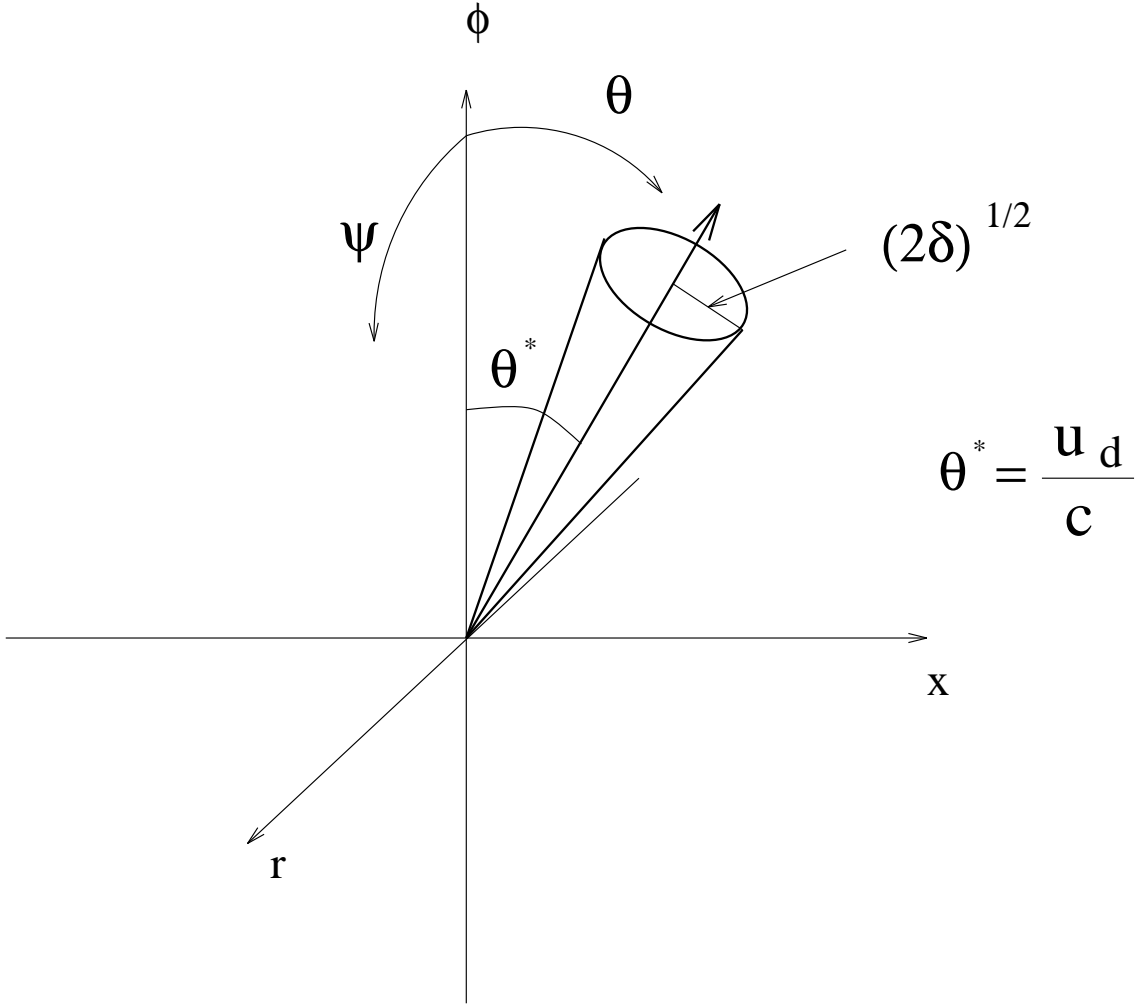,width=15.0cm}
\caption{ 
Emission geometry of the Cherenkov-drift resonance
\label{Cherenkovdrift1}}
\end{figure}

The resonance condition, given by the delta function in Eq. (\ref{qaq7}), then reads
\begin{equation}
{1 \over 2 \,  \gamma_{res}^2 } - \delta +
{ k_r^2  \over 2 k_{\phi}^2 }  +
{1 \over 2} \left( { k_x \over k_{\phi}}  - {u_d\over c } \right)^2 = \,0 ,
\label{aa11}
\end{equation}
where we used $v_{res}=
 c(1- {1 \over 2 \,  \gamma_{res}^2 } - {u_d^2\over 2 c^2}
) $.  The emission geometry at  the Cherenkov-drift resonance is shown in
Figs. \ref{Cherenkov-driftemission}
and \ref{Cherenkovdrift1}.

The maximum growth rate for the t-mode is reached when $k_x/k_{\phi} = u_d/c$
and the maximum growth rate for the lt-mode is reached when $k_r=0$.
We also note, that in the  excitation of both lt- and t-wave it is the 
$x$ component of the electric field that is growing exponentially.

Estimating (\ref{qaq7}) and  (\ref{qaq71}) using $\delta$-function 
( { \rm max } [ ${k_r \over k_{\perp} }$ ] $
\approx c \sqrt{2  \delta } /u_d$ and 
max[$ {k_{\phi} k_x u_d  \over c k k_{\perp}} - { v_{\phi} \over c} { k_{\perp} 
\over k} 
$]  $
\approx \sqrt{  2 \delta} $), 
we wind the maximum growth rates of the t-
and lt-modes
in the limit $ \delta \gg 1/\gamma_{res}^2 $:
\begin{equation}
\Gamma ^{t} =  \Gamma ^{lt}
  \approx  {2 \omega_{p, res}^2  \delta \over  \omega }  
\left(  
{\gamma^3 \over  1+ u_d ^2 \,  \gamma^2 /c^2  } 
{\partial  f(\gamma)  \over \partial  \gamma } \right)_{res} ,
\label{qaq81}
\end{equation}
where $\omega_{p,res}$ is the plasma density of the resonant particles.

We estimate the growth rates (\ref{qaq81}) for the distribution
function of the resonant particles having a Gaussian form:
\begin{equation}
f(p_{\phi} )= { 1 \over \sqrt{2 \pi} p_t } 
 \exp\left( -{ (p_{\phi} -p_b)^2 \over 2  p_t^2}
 \right),
\label{qq1}
\end{equation}
where $p_b$  is the momentum of the bulk motion of the beam and
$p_t$ is the dispersion of the momentum. 
Assuming in (\ref{qaq81}) that $ u_d \gamma_b /c \gg 1$ we find
the growth rates 
\begin{equation}
\Gamma ^{t} =  \Gamma ^{lt}  \approx \sqrt{ { 2\over \pi}} 
  { \omega_{p, res}^2  \delta \gamma_b  \over  \omega  \gamma_t^2 }  
{c^2 \over u_d^2} ,
 \label{qaq82}
\end{equation}
where $\gamma_b = p_b/( mc),\, \gamma_t= p_t/( m c)$.

Numerical estimates show, that the growth rate (\ref{qaq82}) may be large enough
to account for the high brightness radiation emission generation in pulsars.
For the sake of consistency we leave
the detail investigation of the possible application of this radio emission mechanism
to pulsar physics
for a separate paper.

\section{Conclusion}

In this paper we  considered a new  Cherenkov-drift 
emission mechanism that combines
 features of the conventional cyclotron, Cherenkov and curvature emission. 
We argued, that from the microphysical point of view this emission mechanism
may be regarded as a
Cherenkov-type process in inhomogeneous magnetic field. 
Considering emission process in  cylindrical coordinates we have obtained 
 the  single particle emissivities.
We also pointed out, that in order to obtain correct expressions for the emissivities
it is necessary to use the polarization vectors of the normal modes of the medium.
Finally, we calculated the growth rates of the Cherenkov-drift 
instability in a strongly magnetized electron-positron plasma. 

\acknowledgments
We would like to thank George Melikidze and Qinghuan Luo for their
comments.
This research was supported by grant AST-9529170.

\end{document}